\documentclass[a4paper, 12 pt]{article}
\usepackage[utf8]{inputenc}
\usepackage{jheppub}
\pdfoutput=1
\usepackage[english]{babel}
\usepackage{amsmath}
\usepackage{amssymb}
\usepackage{physics}
\usepackage{graphicx}
\usepackage{caption}
\usepackage{subcaption}
\usepackage{hyperref}
\usepackage{esvect}

\title{\vspace{-2cm}
	\begin{flushright}
		{\normalsize INR-TH-2025-016}
	\end{flushright}
	\vspace{0.5cm} Pion bremsstrahlung in the splitting function formalism and the dark photon production}

\author[a,b]{D. Gorbunov,}
\author[a,c]{E. Kriukova}

\affiliation[a]{Institute for Nuclear Research of the Russian Academy of Sciences, \\
	60th October Anniversary pr-ct 7a, Moscow 117312, Russia}
\affiliation[b]{Landau Phystech School of Physics and Research, Moscow Institute of Physics and Technology, \\
	Institutskiy per. 9, Dolgoprudny 141700, Russia}
\affiliation[c]{Institute of Theoretical and Mathematical Physics, Lomonosov Moscow State University,\\
	Lomonosovsky pr-ct 27-1, Moscow 119991, Russia}

\emailAdd{gorby@ms2.inr.ac.ru}
\emailAdd{kriukovaea@my.msu.ru}

\abstract{We study the production of hypothetical vector portal mediators, dark photons $\gamma^\prime$, with masses in the range 0.4--3.5\,GeV in the negatively charged pion-proton collisions $\pi^-p\rightarrow \gamma^\prime X$ via inelastic pion bremsstrahlung and QCD Drell--Yan-like process. In both cases we estimate the value of total dark photon production cross section and obtain the energy distribution for dark photons that could be produced in the NA64h experiment. We also present the mean energies of dark photons produced in the same way by the secondary pions with momenta typical for T2K, DUNE and SHiP experiments.}

\begin{document}
    \maketitle
	\flushbottom
\section{Introduction}
To date, a number of experimental facts and the results of astronomical observations, e.g. neutrino oscillations, galactic rotation curves, CMB measurements, etc. strongly indicate the need to extend the Standard Model (SM). There are many possibilities for such extensions beyond the SM (BSM), so it seems most logical to start with the simplest models with the smallest number of new species and parameters. One of the convenient ways to classify these simplest renormalizable BSM extensions is the portal formalism~\cite{Beacham:2019nyx}. 

In this paper, we consider the phenomenology of the vector portal mediator, dark photon $A^\prime_\mu$ with mass $m_{\gamma^\prime}$~\cite{Okun:1982xi,Galison:1983pa,Holdom:1985ag}. We take the minimal scenario (BC1 in Physics Beyond Colliders classification~\cite{Beacham:2019nyx}) in which a dark photon cannot decay into pairs of dark sector particles, for example, due to kinematic constraints. This results in the following interaction Lagrangian for dark photon, describing its coupling to the SM electromagnetic current $J_\text{em}^\mu$ \cite{Miller:2021ycl}
\begin{equation}
\label{interaction}    
{\cal L}_\text{int}=-\epsilon e J_\text{em}^\mu A^{\prime}_\mu\,, 
\end{equation}
where $\epsilon$ is the kinetic mixing parameter of dark photon and $e$ is the proton electromagnetic charge. Thus, the minimal model is characterized by two parameters, $(m_{\gamma^\prime}, \epsilon)$.

Until recently, it was widely accepted that there exist three key mechanisms of dark photon production in experiments with a proton beam at fixed target: neutral meson decays, e.g. $\pi^0(\eta)\rightarrow \gamma^\prime \gamma$, inelastic proton bremsstrahlung, $pp\rightarrow \gamma^\prime X$, and the analogue of the Drell--Yan process $q\bar{q}\rightarrow \gamma^\prime$~\cite{Kyselov:2024dmi}. Pion bremsstrahlung, as the fourth possible production mechanism, has been considered in~\cite{Curtin:2023bcf} for the first time. In particular, it was claimed that in the SpinQuest experiment at Fermilab~\cite{Berlin:2018pwi}, where protons with momentum $P=120\,\text{GeV}$ hit the iron target, the number of events associated with dark photon production via secondary pion bremsstrahlung exceeds the similar number of events via the Drell--Yan process at least a few times (see figure 5 in~\cite{Curtin:2023bcf}). However, we have several questions about the applicability of chiral perturbation theory (ChPT) in the original paper, which we outline in this paper.

There are two types of experiments for which pion bremsstrahlung should be carefully studied. First of all, the secondary pions are intensively produced in the fixed-target experiments with high-energy proton beams, such as SpinQuest~\cite{Berlin:2018pwi}, T2K~\cite{Araki:2023xgb}, DUNE~\cite{Breitbach:2021gvv}, and SHiP~\cite{SHiP:2020vbd}. As we show below, some of these pions are energetic enough to produce a dark photon of mass 0.4--3\,GeV. In projects with lower energy protons the secondary pions can produce only lighter dark vectors, and this contribution is generically subdominant, e.g. see figure 6 in \cite{Demidov:2025yyr} for the TiMoFey project. 
Second, in the NA64h experiment at CERN SPS, the iron target is hit by the beam of negatively charged pions of momenta $P=50\,\text{GeV}$. Searches for dark photons in neutral meson decays at NA64h have been proposed in~\cite{Gninenko:2023rbf}. The first NA64h results on constraining the branching ratio $\eta(\eta^\prime)\rightarrow\text{invisible}$ are recently presented in~\cite{NA64:2024mah}. Therefore, in this paper, we revisit pion bremsstrahlung as a source of dark photons and develop new approaches to it.

The paper is organized as follows. In section~\ref{sec:ChPT} we discuss the applicability of ChPT for beams of energetic pions. We propose an alternative estimate for the pion bremsstrahlung cross section, focusing on dark photons with masses 0.4--1.3\,GeV, in section~\ref{sec:splitting}. For dark photons with masses 1.3--3\,GeV, the dominant contribution to dark photon production comes from an additional process of Drell--Yan type. Numerical estimates of its cross section are given in section~\ref{sec:qcd}. Finally, section~\ref{sec:discussion} summarizes our findings.

\section{Applicability of ChPT for elastic \texorpdfstring{$\pi^-p$}{pip}-scattering} \label{sec:ChPT}

The authors of \cite{Curtin:2023bcf} described the pion bremsstrahlung in the framework of the leading order Heavy Baryon ChPT. In this work, in general, we adopt  the tree-level Lagrangian from ref.\,\cite{Curtin:2023bcf}. Its part, relevant for elastic $\pi^-p$-scattering, that is discussed below in this section, reads
\begin{equation}
\begin{split}
    \mathcal{L}_\text{int}&=-ieA^\mu\left(\pi^-\partial_\mu\pi^+-\pi^+\partial_\mu\pi^-\right)-eA_\mu \bar{p}\gamma^\mu p \\&-\frac{g_A}{F\sqrt{2}}\left(\bar{p}\gamma^\mu \gamma^5 n \partial_\mu \pi^++\bar{n}\gamma^\mu\gamma^5 p \partial_\mu \pi^-\right)+\frac{i}{4F^2}\bar{p}\gamma^\mu p \left(\pi^+\partial_\mu \pi^- - \pi^-\partial_\mu \pi^+\right),
\end{split}
\end{equation}
where the interesting particles are proton $p$, neutron $n$, photon $A_\mu$ and pions $\pi^\pm$, and the model parameters are the axial coupling $g_A=1.27$ and the pion decay constant $F=93\,\text{MeV}$.
The only change that we have made in accordance with several works specializing in ChPT~\cite{Hemmert:1997ye,Fettes:2000gb,Rijneveen:2021bfw} is the change of sign in the definition of $u_\mu\equiv i u^\dagger (\nabla_\mu U) u^\dagger$. Effectively, this flips the sign of the axial coupling $g_A \rightarrow -g_A$, but the squared matrix elements at the tree level are not affected at all.

Here we should stress that ChPT is the low-energy effective field theory and thus at the tree level it can be safely applied only to small values of the expansion parameter $Q/(4\pi F)$, where $Q$ is the characteristic momentum scale of the process or i.e. the center-of-mass energy~\cite{Coogan:2021sjs}. When the expansion parameter becomes comparable to 1, both loop diagrams based on the leading order (LO) ChPT Lagrangian, and tree-level diagrams, based on the next-to-leading order (NLO) ChPT Lagrangian, give considerable input in comparison with the tree-level LO contributions, so one cannot ignore them any more in the calculations. 

Moreover, even for lower momenta $Q\gtrsim600\,\text{MeV}$ the corrections from vector mesons should be taken into account~\cite{Ecker:1988te}. For instance, consider the example of $\rho$-meson. By adding it to the ChPT as a dynamical vector degree of freedom, one obtains the following interactions of $\rho^0$-meson with the charged pions
\begin{equation}
    \mathcal{L}_{\rho \pi\pi}=\frac{ig_{\rho \pi \pi}\sqrt{2}}{F^2} \partial_{\left[\mu\right.}\rho^0_{\left.\nu\right]} \partial^{\left[\mu\right.}\pi^-\partial^{\left.\nu\right]}\pi^+
\end{equation}
and with protons
\begin{equation}
    \mathcal{L}_{\rho pp}=g_{\rho NN} \bar{p}\gamma^\mu \rho^0_\mu p,
\end{equation}
where $g_{\rho \pi \pi}\simeq0.05$ and $g_{\rho NN}\simeq 2.0$~\cite{Ecker:1988te,Ecker:1989yg,Schindler:2005ke}. Besides, it is well-known that charged pions and protons interaction with $\Delta$-resonances has a significant impact on $\pi^-p$-scattering amplitudes~\cite{Scherer:2012xha} for momenta $m_\Delta\gtrsim Q\gtrsim1\,\text{GeV}$. At the tree level, $\Delta$-resonances can be included in the process under consideration using the Lagrangian
\begin{equation}
    \mathcal{L}_{\Delta \pi p}=\frac{g_{\Delta\pi p}}{F}\left(\overline{\Delta^{++}_\mu}\Theta^{\mu\nu}\partial_\nu\pi^+ p-\frac{1}{\sqrt{3}}\overline{\Delta^0_\mu}\Theta^{\mu\nu} \partial_\nu\pi^-p\right)+h.c.,
\end{equation}
where $g_{\Delta \pi p}\simeq 1.25$, $\Theta^{\mu\nu}\equiv g^{\mu\nu}+z_0\gamma^\mu\gamma^\nu$, and $z_0\simeq -0.22$~\cite{Vonk:2022tho}.

Let us now estimate the values of typical center-of-mass energies of the colliding negatively charged pion with the four-momentum $p_1$ and mass $m_\pi$ and the proton with the four-momentum $p_2$ and mass $M_p$. To make a numerical estimate we take the pion energy 
$P=50\,\text{GeV}$, expected to be used in one of the perspective experiments, NA64h. Given the four-momenta of colliding particles in the laboratory frame
\begin{equation}
    p^\text{lab}_1=\{E^\text{lab}_1, 0, 0, P\}, \quad p^\text{lab}_2=\{M_p,0,0,0\},
\end{equation}
we construct the Mandelstam variable
\begin{equation}
    s = m^2_\pi + 2E^\text{lab}_1M_p+M^2_p
\end{equation}
and the center-of-mass energies
\begin{equation}
    E^\text{cm}_1=\frac{m^2_\pi+M_pE^\text{lab}_1}{\sqrt{s}}=4.8\,\text{GeV}, \quad E^\text{cm}_2=\frac{M^2_p+M_pE^\text{lab}_1}{\sqrt{s}}=4.9\,\text{GeV},
\end{equation}
which are definitely outside the region of the LO ChPT applicability.

And yet ChPT has been applied in~\cite{Curtin:2023bcf} to describe the process of pion brems\-strah\-lung for pion beams of even higher energies. The authors justified its validity by imposing the cut on the negative square of the four-momentum transferred to the proton, $t<(4\pi F)^2$. This procedure seems questionable to us, and this is why below we study the validity of the results obtained within the leading order ChPT for high-energy pion beams. In order to make the arguments as clear as possible, we consider the simple example of $\pi^-p$ elastic scattering. 

We keep the notation for the four-momenta of the incoming pion $p_1$ and the target proton $p_2$ and denote the four-momenta of the incident pion and the proton in the final state as $p_3$ and $p_4$ correspondingly. We also introduce the angle $\theta$ between the proton three-momenta $\vec{p}^\text{\,\,cm}_2$ and $\vec{p}^\text{\,\,cm}_4$ in the center-of-mass frame.
Then, in the laboratory frame the Lorentz transformation gives the energy of proton in the final state 
\begin{equation}
    E^\text{lab}_4=\gamma E^\text{cm}_2 - \gamma\beta p^\text{cm}_1\cos\theta,
\end{equation}
where $p^\text{cm}_1\equiv\sqrt{\left(E^\text{cm}_1\right)^2-m^2_\pi}$ and the Lorentz transformation parameters used to switch to the center-of-mass frame from the laboratory frame read
\begin{equation}
    \gamma=\frac{M_p+E^\text{lab}_1}{\sqrt{s}}, \quad \gamma\beta=\frac{P}{\sqrt{s}}.
\end{equation}
Imposing the cut on the energy transferred to proton
\begin{equation}
	E^\text{lab}_4-M_p \leq 4\pi xF,
\end{equation}
where $x$ is the auxiliary dimensionless parameter that we can vary between 0 and 1, we obtain the following constraint on the angle $\theta$, ensuring the naive ``validity'' of ChPT in the same manner as it was made in~\cite{Curtin:2023bcf},
\begin{equation}
	\cos\theta \geq \frac{\left(M_p+E^\text{lab}_1\right)E^\text{cm}_2-(M_p+4\pi xF)\sqrt{s}}{p^\text{cm}_1 P}.
\end{equation}
For the pion beam of the NA64h experiment with $P=50\text{\,GeV}$, this gives $\cos\theta \geq 0.995$ for $x=0.1$ and $\cos\theta \geq 0.95$ for $x=1$. 

\begin{figure}[!htb]
	\begin{center}
		\begin{subfigure}{0.31\textwidth}
			\centering
                \includegraphics[width=\textwidth]{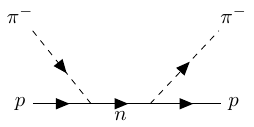}
			\caption{}
            \label{fig:pipsch}
		\end{subfigure}
        \hfill
		\begin{subfigure}{0.31\textwidth}
			\centering
			\includegraphics[width=\textwidth]{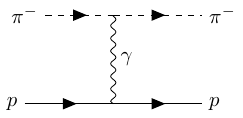}
			\caption{}
            \label{fig:piptch}
		\end{subfigure}
        \begin{subfigure}{0.31\textwidth}
			\centering
			\includegraphics[width=\textwidth]{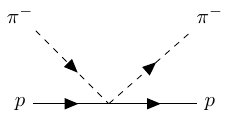}
			\caption{}
		\end{subfigure}
	\end{center}
	\caption{Feynman diagrams for elastic $\pi^-p$-scattering in the leading order of Heavy Baryon ChPT: (a) $s$-channel, (b) $t$-channel, (c) 4-point. Figures are made with the help of the package \texttt{TikZ-Feynman}~\cite{Ellis:2016jkw}.}
        \label{fig:pipFeyn}
\end{figure}

Figure~\ref{fig:pipFeyn}
shows the Feynman diagrams for the process $\pi^-p\rightarrow \pi^-p$ in the leading order of ChPT. The corresponding matrix elements were calculated and squared using the package \texttt{CalcHEP}~\cite{Belyaev:2012qa}. We have checked that (just as in the process of pion bremsstrahlung, which was studied in~\cite{Curtin:2023bcf}) for $\pi^-p$ elastic scattering the diagram with $t$-channel photon exchange makes a negligible contribution compared to the two other diagrams. This means that, unlike the classical textbook example of electron-proton scattering, the amplitudes of $\pi^-p$-scattering and of the elastic pion bremsstrahlung $\pi^-p\rightarrow\pi^-p\gamma^\prime$ are not determined solely by the Mandelstam variable $t$. Therefore, we conclude that for $\pi^-p$-scattering as well as for the similar process of elastic pion bremsstrahlung it is incorrect to justify the applicability of the leading order of ChPT based only on the smallness of the square of the momentum transferred to the proton or, equivalently, the smallness of the transferred energy or the proximity to zero of the scattering angle of the proton.

Except those shown in figure~\ref{fig:pipFeyn}, there are additional diagrams with had\-ro\-nic resonances at the tree level, for instance, the scattering in the $s$-channel with $\Delta^0$-baryon (the analogue of figure~\ref{fig:pipsch}), the $\rho$-meson exchange in the $t$-channel (the analogue of figure~\ref{fig:piptch}), and $\pi^-p$-scattering in the $u$-channel with $\Delta^{++}$-baryon. At first glance, it seems that for a more accurate analysis of $\pi^-p$-scattering in ChPT, one needs to take into account all these diagrams, as well as the NLO contributions. However, upon closer examination, it becomes evident that in order to consider the pion and proton beams with the center-of-mass energies $E^\text{cm}_1\simeq E^\text{cm}_2 \simeq 5\,\text{GeV}$, it is  necessary to take into account the impacts of all possible resonances with masses up to 5 GeV, otherwise the answer will obviously be incomplete. It is clear that this cannot be done within the framework of ChPT, or even just the perturbative analytical approach, so we limit ourselves to the LO calculation with only pions and nucleons and refer the reader to the comment on a possible diffraction physics approach at the end of this section.

\begin{figure}[!tb]
    \centering
    \includegraphics[width=0.6\linewidth]{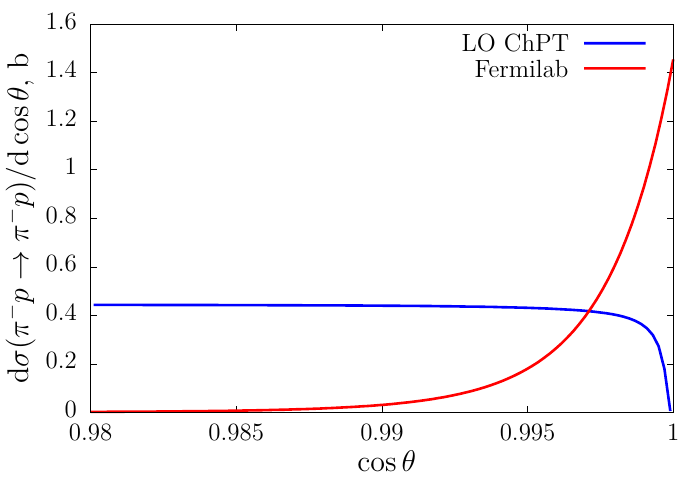}
    \caption{Differential cross section of elastic pion-proton scattering $\pi^-p\rightarrow\pi^-p$ as the function of $\cos\theta$, where $\theta$ is the proton scattering angle in the center-of-mass frame and the momentum of the incident pion in the lab frame $P=50\text{\,GeV}$. The results of the LO ChPT calculation are shown in blue. The fit of Fermilab experimental data~\cite{FermilabSingleArmSpectrometerGroup:1976krf} for the pion beam of the same momentum is presented in red for comparison.}
    \label{fig:pipScat}
\end{figure}

In figure~\ref{fig:pipScat}
we show in blue the differential cross section of elastic $\pi^-p$-scattering as the function of $\cos\theta$ obtained in the framework of LO ChPT using \texttt{CalcHEP} for the pion momentum in the lab frame $P=50\text{\,GeV}$. This process for the same pion beam momentum has been studied at the Single Arm Spectrometer Facility in Fermilab~\cite{FermilabSingleArmSpectrometerGroup:1976krf}. 
In particular, it was shown that the angular distribution for the differential cross section of elastic $\pi^-p$-scattering can be described adopting the fit
\begin{equation}
    \dd \sigma/\dd t=A\,e^{B\abs{t}+C\abs{t}^2} 
\end{equation}
with the parameters $A=31.3\text{\,mb\,GeV}^{-2}$, $B=-9.7\text{\,GeV}^{-2}$ and $C=3.1\text{\,GeV}^{-4}$. In the center-of-mass frame the Mandelstam variable equals $t=-2\left(p_1^\text{cm}\right)^2(1-\cos\theta)$. It allows for changing variables and obtaining the Fermilab fit for $\dd\sigma/\dd\cos\theta$, which is shown in red in figure~\ref{fig:pipScat}. One concludes from the plot of figure~\ref{fig:pipScat} that LO ChPT cannot be applied to describe the elastic pion-proton scattering at the energies typical for NA64h. Furthermore, imposing cuts on $\cos \theta$ does not improve the situation in any sense. 

Thus, we conclude that the use of such cuts for the pion bremsstrahlung process, which is described by the same diagrams of figure~\ref{fig:pipFeyn} with the one outgoing dark photon line attached, is also likely to be unhelpful, and the LO ChPT simply cannot be applied at such energies. Therefore, it is necessary to look for new methods for calculating the cross section of the pion bremsstrahlung. In the next section, we propose a method based on factorization and utilizing numerical fits to experimental data.

Let us note that the method presented below estimates the \textit{inelastic} pion brems\-strah\-lung cross section. The study of elastic pion brems\-strah\-lung lies outside the scope of this work, but nevertheless can be an interesting problem. The elastic $\pi p$-scattering was earlier successfully described in the Regge regime of the holographic QCD model~\cite{Liu:2023tjr}. In contrast to ChPT, taking into account the Pomeron and Reggeon exchanges allows to correctly predict the differential cross section of elastic $\pi p$-scattering. It seems that the elastic pion bremsstrahlung can also be considered either in the extended holographical QCD model with the addition of electromagnetic interaction, or by accounting for the non-zero momentum transfer in analogy with the elastic proton bremsstrahlung~\cite{Gorbunov:2023jnx}.

\section{Splitting function for pion bremsstrahlung} \label{sec:splitting}

In this section, we present the new approach to estimate the cross section of hidden photon production via inelastic pion bremsstrahlung, $\pi^-p\rightarrow \gamma^\prime X$. It is largely based on the factorization procedure, originally implemented by Altarelli and Parisi~\cite{Altarelli:1977zs} for partons. Later, the idea of factorization was intensively used to describe the production of dark massive scalars and vectors with masses about $\mathcal{O}(1)\text{\,GeV}$ in the process of inelastic proton bremsstrahlung~\cite{Boiarska:2019jym,Foroughi-Abari:2021zbm,Foroughi-Abari:2024xlj,Gorbunov:2024iyu,Gorbunov:2024vrc} (see also~\cite{Alimena:2025kjv} for a brief review on key results). The applicability of this method to the initial state radiation for proton bremsstrahlung was verified in~\cite{Foroughi-Abari:2024xlj} by comparing the estimated cross section with the experimental data on the inclusive $\rho(770)$-meson production in $pp$-collisions. 

\begin{figure}
    \centering
    \includegraphics[width=0.35\linewidth]{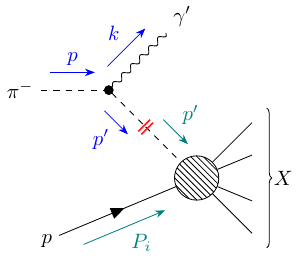}
    \caption{Feynman diagram for the dark photon production in the initial state radiation of inelastic pion bremsstrahlung.}
    \label{fig:pionBremDiag}
\end{figure}

Figure~\ref{fig:pionBremDiag}
shows the Feynman diagram for dark photon production in the inelastic pion bremsstrahlung $\pi^-p\rightarrow\gamma^\prime X$. Below we denote the 4-momenta of incident pion, dark photon, target proton, and intermediate virtual pion as $p$, $k$, $P_i$, and $p^\prime\equiv p-k$ correspondingly. Thus, in the laboratory frame by choosing the $z$-axis direction along the incident pion's momentum, we obtain the following components of four-momenta: 
\begin{align}
    p=& \,\{E_\pi, 0, 0, P\},\\
    k=& \,\{E_{\gamma^\prime}, k_\perp\cos\varphi,  k_\perp\sin\varphi, zP\},\\
    P_i=& \,\{M_p, 0, 0, 0\},
\end{align}
where $z$ is the fraction of the longitudinal component of the three-momentum that is taken by the dark photon from the incident pion, $k_\perp$ is the transverse part of the dark photon 3-momentum, and for a high-energy pion beam we approximate the incident pion energy as $E_\pi\simeq P+m^2_\pi/(2P)$ and the dark photon energy as $E_{\gamma^\prime}\simeq zP+\left(m^2_{\gamma^\prime}+k^2_\perp\right)/\left(2zP\right)$. In what follows, we consider only the impact of dark photons with kinematic variables obeying 
\begin{equation}
\label{con-1}    
\frac{m^2_{\gamma^\prime}+k^2_\perp}{z^2P^2}\leq0.1\,.
\end{equation}

The general idea of our calculation is to factorize the input of the subprocess $\pi^-(p)\rightarrow \gamma^\prime(k) \pi^-(p^\prime)$ in terms of the so-called splitting function $w_\pi(z, k^2_\perp)$ and to connect the cross section of inelastic pion bremsstrahlung $\pi^-p\rightarrow\gamma^\prime X$ with the cross section of inelastic process $\pi^-p\rightarrow X$. Therefore, we should first focus on the charged pion electromagnetic vertex $\pi\pi\gamma$. For both initial and final pions being on-shell, the matrix element of the electromagnetic current $J^\mu_\text{em}=\sum_{q=u,d} Q_q e\bar{q}\gamma^\mu q$ (with proton charge $e$ and quark charges $Q_qe$) reads \cite{RuizArriola:2024gwb}
\begin{equation}
    \bra{\pi^-(p^\prime)}J^\mu_\text{em}(0)\ket{\pi^-(p)}=-eF^\pi_\text{em}((p^\prime-p)^2)(p+p^\prime)^\mu\equiv -e\Gamma^\mu_0,
\end{equation}
where $F^\pi_\text{em}$ is the charged pion electromagnetic form factor, and $\Gamma^\mu_0$ is the on-shell vertex function. If the initial pion is on-shell, $p^2=m^2_\pi$, but the final pion is not, $p^{\prime2}\equiv t\neq m_\pi^2$, the generalized half-off-shell vertex function $\Gamma^\mu_{1/2}$ can be expressed with the help of the Ward--Takahashi identity as \cite{Rudy:1994qb,Choi:2019nvk,Leao:2024agy}
\begin{equation}
    \Gamma^\mu_{1/2} = (p+p^\prime)^\mu F_1(Q^2, t) + (p-p^\prime)^\mu \frac{(t-m^2_\pi)}{Q^2}\left(F_1\left(0,t\right)-F_1\left(Q^2, t\right)\right),
\end{equation}
where the half-off-shell pion electromagnetic form factor $F_1$ depends on the negative square of the momentum transfer $Q^2\equiv -(p^\prime-p)^2$ and on $t$. In our case, the second term does not contribute to the final answer. Indeed,  the expression for the amplitude of pion bremsstrahlung contains the dark photon polarization vector orthogonal to $(p-p^\prime)_\mu$, so we keep only the first term for the vertex function.

Following the logic applied in the works on proton bremsstrahlung~\cite{Foroughi-Abari:2021zbm,Foroughi-Abari:2024xlj,Gorbunov:2024iyu,Gorbunov:2024vrc}, we introduce the phenomenological off-shell hadronic form factor~\cite{Feuster:1997pq}
\begin{equation}
    F_\text{virt}\equiv \frac{\Lambda^4}{\Lambda^4+H_\pi^2\left(z, k^2_\perp\right)/z^2},
\end{equation}
depending on the energy scale $\Lambda = 1.2\text{ GeV}$ and on the kinematic combination 
\begin{equation*}
H_\pi \equiv m^2_\pi z^2 + m^2_{\gamma^\prime}(1-z)+k^2_\perp\,. 
\end{equation*}
Next, we present the half-off-shell form factor as the product 
\[
F_1(Q^2, t)\simeq F^\pi_\text{em}((p^\prime-p)^2) F_\text{virt}(z, k^2_\perp)\,.
\]
In the numerical calculations below, we adopt the value of the pion electromagnetic form factor $F^\pi_\text{em}$ obtained in~\cite{RuizArriola:2024gwb} utilizing BaBar data~\cite{BaBar:2012bdw} and dispersion relations. 

Applying the standard Feynman diagram technique, we factorize the inelastic pion brems\-strahlung amplitude
\begin{equation} \label{eq:matrelem}
    \mathcal{M}^{r\lambda}_{\pi p\rightarrow \gamma^\prime X}=\mathcal{M}^r_{\pi p\rightarrow X}\frac{1}{p^{\prime 2}-m^2_\pi}(-\epsilon e) F^\pi_\text{em}(m^2_{\gamma^\prime}) F_\text{virt}(z, k^2_\perp) (p+p^\prime)_\mu (\epsilon^\lambda)^{*\mu},
\end{equation}
i.e. express it through the amplitude of the inelastic $\pi^-p$-scattering $\mathcal{M}^r_{\pi p\rightarrow X}$, where $r$ is the spinor index of the target proton and $\lambda$ is the dark photon polarization ($\lambda=\overline{1,3}$). 

Until now, we have calculated matrix elements in the framework of the modern Feynman diagram technique. As we have already mentioned, in this technique the 4-momentum is conserved in every vertex, so $p^\prime\equiv p-k$. It is important to mention that the significant number of works on the proton bremsstrahlung of dark photon, e.g.~\cite{Foroughi-Abari:2021zbm,Foroughi-Abari:2024xlj,Kling:2025udr}, and of dark scalar~\cite{Boiarska:2019jym} use the alternative, \textit{time-ordered} perturbation theory. It assumes that all particles in the vertices are on-shell and that 3-momenta are conserved in every vertex. For inelastic pion bremsstrahlung, these conditions inevitably lead to the non-conservation of the energy in the $\pi\pi\gamma^\prime$ vertex, making the approach non-covariant. Thus, in the \textit{quasi-real} approach~\cite{Kessler:1960,Baier:1973ms,Foroughi-Abari:2024xlj} the components of the intermediate pion 4-momentum are 
\begin{equation}
    p^\prime=\{E^\prime_\pi, -k_\perp\cos\varphi, -k_\perp \sin \varphi, (1-z)P\},
\end{equation}
where the intermediate pion energy is approximated as
\begin{equation}
E^\prime_\pi\simeq (1-z)P + \frac{m^2_\pi+k^2_\perp}{2(1-z)P}.
\end{equation}
The validity of the Taylor expansion used above is provided for processes with momenta obeying the condition 
\begin{equation}
    \label{con-2}
\left(m^2_\pi+k^2_\perp\right)/((1-z)^2P^2)\leq0.1\,.
\end{equation} 
The equivalent and more convenient way to present the intermediate pion 4-mo\-men\-tum in the on-shell approach is to introduce the auxiliary 4-momentum $\delta\equiv \{\Delta E, \vec{0}\}$ with the only non-zero component corresponding to the non-conservation of energy in $\pi\pi\gamma^\prime$-vertex $\Delta E\equiv E^\prime_\pi+E_{\gamma^\prime}-E_\pi\simeq H_\pi/\left(2z(1-z)P\right)$, so that $p^\prime=p-k+\delta$~\cite{Foroughi-Abari:2024xlj}. In what follows, we keep $\delta$ in the calculations. In order to reproduce the results according to modern covariant Feynman technique, one simply needs to replace $\delta$ with zero. Of course, we expect that the final result should not depend on whether we have chosen to work within Feynman diagram technique or time-ordered perturbation theory. 

As an example, consider the intermediate pion propagator. In the usual modern covariant technique, we obtain $\left(p^{\prime 2}-m^2_\pi\right)^{-1}=-z/H_\pi$. The time-ordered perturbation theory states that there are two diagrams, depending on the time order of processes: (\textit{i}) either at first happens the dark photon emission $\pi^-\rightarrow\pi^-\gamma^\prime$ and then the pion-proton inelastic interaction $\pi^-p\rightarrow X$, or (\textit{ii}) inelastic positive pion emission $p\rightarrow \pi^+ X$ with the subsequent annihilation $\pi^-\pi^+\rightarrow \gamma^\prime$. In the quasi-real approximation, one neglects the impact of process (\textit{ii}) in comparison with process (\textit{i}), given that $|E_\pi-E_{\gamma^\prime}-E^\prime_\pi|\ll E_\pi+E^\prime_\pi-E_{\gamma^\prime}$, i.e. 
\begin{equation}
\label{con-3}
H_\pi/z \ll \left(2(1-z)P\right)^2\,. 
\end{equation}
Below we check that this condition holds in the kinematic region considered in the numerical calculations. The impact from the (\textit{i}) diagram to the pion ``propagator'' turns out to be the same, as in the modern QFT approach, $\left(2E^\prime_\pi\left(E_\pi-E_{\gamma^\prime}-E^\prime_\pi\right)\right)^{-1}\simeq-z/H_\pi$~\cite{Baier:1973ms,Boiarska:2019jym,Foroughi-Abari:2024xlj}.

To obtain the squared matrix element from~\eqref{eq:matrelem}, we first need to calculate
\begin{equation} \label{eq:W}
    \mathcal{W}\equiv \left(2p-k+\delta\right)_\mu \left(2p-k+\delta\right)_\nu \sum_\lambda \left(\epsilon^\lambda\right)^{*\mu} \left(\epsilon^\lambda\right)^\nu.
\end{equation}
We start from an illustrative example of the massless vector boson, e.g. SM photon, that has only two transverse polarizations. This result can be considered as an effective massless limit $m_{\gamma^\prime}\rightarrow0$ for a dark photon. As it was discussed in \cite{Baier:1973ms} for electron bremsstrahlung, in order to obtain the self-consistent answer in the quasi-real approximation, one should choose the polarization 4-vectors as a space-like vector of the form $\left(\epsilon^{\pm}\right)^\mu=\{0, \vv{\epsilon^\pm}\}$, so that the sum over transverse polarizations $\Sigma^T_{\mu\nu}\equiv \sum_{\lambda=\pm} \left(\epsilon^\lambda\right)^{*\mu} \left(\epsilon^\lambda\right)^\nu$ is non-zero only if both $\mu, \nu \neq 0$. Thus, the non-conservation of energy $\Delta E$ \textit{never} enters the final answer and it is not important, whether we adopt the Feynman technique or non-covariant quasi-real approximation for the calculations. The sum over transverse polarizations can be written as \cite{Greiner:1996zu,Dreiner:2008tw}
\begin{equation}
    \Sigma^T_{\mu\nu}=-g_{\mu\nu}+\frac{k_\mu \bar{k}_\nu+\bar{k}_\mu k_\nu}{k\cdot \bar{k}},
\end{equation}
where $k^\mu=\{E_\gamma,\vec{k}\}$, $\bar{k}^\mu=\{E_{\gamma},-\vec{k}\}$ and we consider the massless particle, so $k^2=0$. It is straightforward to check that $k^\mu \Sigma^T_{\mu\nu}=0$ and $\Sigma_{0\mu}=\Sigma_{\mu 0}=0$ for any $\mu$. In the transverse massless limit, this gives the following result for \eqref{eq:W}
\begin{equation}
\label{W1}
    \mathcal{W}^T=\frac{4k^2_\perp}{z^2}.
\end{equation}

Having in mind the above calculation, we intend to find the analogous choice for the sum over massive boson polarizations. It can be easily seen that the naive sum over polarizations from a textbook, $-g_{\mu\nu}+k_\mu k_\nu/m^2_{\gamma^\prime}$, does not fulfil the quasi-real approximation requirement for polarization sum, i.e. it does not vanish if $\mu$ or $\nu$ is zero. Moreover, it is evident that the production of extremely high-energetic massive vector boson can potentially lead to unitarity violation. As is well known from the investigations of the electroweak sector of the Standard Model, the unitarity gets restored when taking into account diagrams with (dark) Higgs. In other words, the contribution of the Goldstone mode to the longitudinal polarization of the vector boson should ultimately be reduced by a similar contribution from the (dark) Higgs boson, which was not explicitly included in our effective analysis. It may be justified in the models, in which dark Higgs boson is much heavier than the dark photon and thus the dark Higgs production in scatterings is kinematically forbidden. Otherwise the unitarity must be restored effectively. To this end, we apply the Dawson correction, earlier used in \cite{Foroughi-Abari:2024xlj,Kling:2025udr}, to subtract the Goldstone mode from the longitudinal po\-la\-ri\-za\-tion vector $\epsilon^\mu_L$ by hands
\begin{equation}
    \epsilon^\mu_0 \equiv \epsilon^\mu_L-\frac{k^\mu}{m_{\gamma^\prime}}=\frac{m_{\gamma^\prime}}{E_{\gamma^\prime}+|\vec{k}|}\left\{-1, \frac{\vec{k}}{|\vec{k}|}\right\}.
\end{equation}
Note that the longitudinal component of the dark vector does not couple to the SM sector, if the electromagnetic current entering the interaction \eqref{interaction} is conserved.
This brings us to the corrected polarization sum 
\begin{equation}
    \Sigma^{\mu\nu}\equiv \sum_{\lambda=\pm} \left(\epsilon_\lambda\right)^{*\mu} \epsilon_\lambda^\nu + \left(\epsilon_0\right)^{*\mu}\epsilon_0^\nu=
    -g^{\mu\nu}-\frac{k^\mu\epsilon^\nu_{0}+k^\nu\epsilon^\mu_0}{m_{\gamma^\prime}}.
\end{equation}
Despite the fact that $\Sigma^{0\mu}= \mathcal{O}\left(H_\pi/\left(z^2 P^2\right)\right)\neq 0$, its impact on the final answer is much smaller than from the components $\Sigma^{ij}=\mathcal{O}(1)$. It is this hierarchy that allows one to apply the quasi-real approximation similar to the massless case. 

Indeed, the decomposition of~\eqref{eq:W} in powers of $\delta$ takes the form
\begin{align}
    \mathcal{W}&\equiv \mathcal{W}_0+2\mathcal{W}_1+\mathcal{W}_2, \label{eq:Wsum} \\
    \mathcal{W}_0 &\equiv \left(2p-k\right)_\mu \left(2p-k\right)_\nu \Sigma^{\mu\nu} = \frac{4H_\pi}{z^2}+m^2_{\gamma^\prime}-4m^2_\pi, \\
    \mathcal{W}_1 &\equiv \left(2p-k\right)_\mu \delta_\nu \Sigma^{\mu\nu}=\mathcal{O}\left(\frac{H^2_\pi}{z^3P^2}\right), \\
    \mathcal{W}_2 &\equiv \delta_\mu \delta_\nu \Sigma^{\mu\nu}=\mathcal{O}\left(\frac{H^3_\pi}{z^4P^4}\right).
\end{align}
Since $H_\pi/\left(zP^2\right)<H_\pi/\left(z^2P^2\right)\ll 1,$ we find $\mathcal{W}_0\gg\mathcal{W}_1\gg\mathcal{W}_2$, hence the latter two terms in~\eqref{eq:Wsum}, proportional to $\delta_\mu$ and $\delta_\mu\delta_\nu$, can be safely omitted. Thus, to the leading order in our approximation we obtain the same result in the quasi-real approximation and using modern covariant Feynman technique, both giving
\begin{equation}
\label{W2}
    \mathcal{W} = \frac{4H_\pi}{z^2}+m^2_{\gamma^\prime}-4m^2_\pi.
\end{equation}
One can also check that $\mathcal{W}\rightarrow \mathcal{W}_T$ in the massless limit $m_{\gamma^\prime}\ll k^2_\perp, m^2_\pi$, see eqs.\,\eqref{W1} and \eqref{W2}. 
The averaged and squared matrix element of pion bremsstrahlung then reads
\begin{equation}
\begin{split}
    \left< |\mathcal{M}|^2\right> \equiv&\,\,\frac{1}{2} \sum_{r, \lambda} |\mathcal{M}^{r \lambda}_{\pi p\rightarrow \gamma^\prime X}|^2=\\=&\sum_r |\mathcal{M}^r_{\pi p\rightarrow X}|^2\frac{\epsilon^2 e^2 z^2}{2H_\pi^2}|F^\pi_\text{em}(m^2_{\gamma^\prime})|^2 F^2_\text{virt}(z, k^2_\perp) \mathcal{W}.
\end{split}
\end{equation}

Next, we proceed with the factorization of the inelastic pion brems\-strah\-lung differential cross section, 
\begin{equation}
    \dd \sigma_{\pi p\rightarrow \gamma^\prime X} = \frac{\left< |\mathcal{M}|^2\right>}{4M_pP} \dd \Phi_{\gamma^\prime X}\,,
\end{equation}
where 
\begin{equation}
    \dd \Phi_{\gamma^\prime X} \equiv \frac{\dd^3 k}{\left(2\pi\right)^3 2E_{\gamma^\prime}} \prod_X \frac{\dd^3 p_X}{\left(2\pi\right)^3 2E_X} \left(2\pi\right)^4 \delta^{(4)}\left(p+P_i-k-\sum_X p_X\right)
\end{equation}
is the inclusive phase space volume with dark photon.  

As a result of straightforward calculation, the differential cross section takes the form of
\begin{equation} \label{eq:crsecfact}
    \frac{\dd^2 \sigma_{\pi p\rightarrow \gamma^\prime X}}{\dd z \dd k^2_\perp} = w_\pi(z, k^2_\perp) |F^\pi_\text{em}(m^2_{\gamma^\prime})|^2 F^2_\text{virt}(z, k^2_\perp) \sigma_{\pi p \rightarrow X}(\bar{s}).
\end{equation}
Here we have introduced the pion splitting function, i.e. the probability of dark photon emission by the incident pion in the subprocess $\pi^-(p)\rightarrow \gamma^\prime(k) \pi^-(p^\prime)$
\begin{equation} \label{eq:splitfunc}
    w_\pi(z, k^2_\perp) = \frac{\epsilon^2 \alpha_\text{em} (1-z)}{4\pi H_\pi} \left(\frac{4}{z}+\frac{z}{H_\pi}\left(m^2_{\gamma^\prime}-4m^2_\pi\right)\right),
\end{equation}
containing the fine-structure constant $\alpha_\text{em}$. In addition, in\,\eqref{eq:crsecfact} we have explicitly extracted the full cross section of the inelastic $\pi^-p$-scattering  
\begin{equation}
    \sigma_{\pi p \rightarrow X} = \frac{1}{4M_pP(1-z)} \int \frac{1}{2} \sum_r |\mathcal{M}^r_{\pi p \rightarrow X}|^2 \dd \Phi_X,
\end{equation}
integrated over the phase space of all the particles possible in the final state
\begin{equation}
    \dd \Phi_X \equiv \prod_X \frac{\dd^3 p_X}{\left(2\pi\right)^3 2E_X} \left(2\pi\right)^4 \delta^{(4)}\left(p^\prime+P_i-\sum_X p_X\right).
\end{equation}
The factorization procedure assumes that the intermediate pion of the momentum $p^\prime\equiv p-k$ is nearly on-shell, which is guaranteed by the suppressing hadronic off-shell form factor $F_\text{virt}(z, k^2_\perp)$. Thus the full inelastic cross section of $\pi^-p$-scattering in\,\eqref{eq:crsecfact} is taken at the Mandelstam variable $s$ equal to $\bar{s}\equiv(P_i+p-k)^2$, or more explicitly: 
\begin{equation}
    \bar{s}=2M_pP(1-z)+M_p^2-\frac{H_\pi}{z}-\frac{M_p}{zP}(m^2_{\gamma^\prime}+k^2_\perp).
\end{equation}

In the numerical calculations, we use the following fit for the total $\pi^-p$-scattering cross section at $\sqrt{s}\geq5\text{\,GeV}$ \cite{ParticleDataGroup:2016lqr}, 
\begin{equation} \label{eq:pipcrsecfit}
    \sigma^{\pi^-p}(s)=H_0\ln^2\left(\frac{s}{s^{\pi p}_M}\right)+P^{\pi p}+R_1^{\pi p}\left(\frac{s}{s^{\pi p}_M}\right)^{-\eta_1}+R_2^{\pi p}\left(\frac{s}{s^{\pi p}_M}\right)^{-\eta_2},
\end{equation}
where $s^{\pi p}_M\equiv\left(m_\pi+M_p+M_0\right)^2$, $M_0=2.1206\text{\,GeV}$ and the values of all the other parameters are listed in table~\ref{tab:sigmapipfit}.
\begin{table}
    \centering
    \begin{tabular}{|c|c||c|c||c|c|}
        \hline
        $H_0,\text{\,mb}$ & 0.2720 & $R_1^{\pi p},\text{\,mb}$ & 9.56 & $\eta_1$ & 0.4473\\
        \hline
        $P^{\pi p},\text{\,mb}$ & 18.75 & $R_2^{\pi p},\text{\,mb}$ & 1.767 & $\eta_2$ & 0.5486\\
        \hline
    \end{tabular}
    \caption{Fit parameters for the total cross section of pion proton scattering $\sigma^{\pi^-p}(s)$ \eqref{eq:pipcrsecfit} at $\sqrt{s}\geq5\text{\,GeV}$~\cite{ParticleDataGroup:2016lqr}.}
    \label{tab:sigmapipfit}
\end{table}
The fit~\eqref{eq:pipcrsecfit} is not valid at $\sqrt{s}<5\text{\,GeV}$, therefore in this region we use the interpolation of tabulated PDG data on total $\pi^-p$-collision cross section \cite{ParticleDataGroup:2022pth}. 

In order to present the numerical results in a more convenient way from the experimental point of view, we change the variables in the differential cross section~\eqref{eq:crsecfact} to the dark photon energy $E_{\gamma^\prime}$ and the polar angle of dark photon 3-momentum $\theta\equiv \arctan\left(k_\perp/(zP)\right)$ using the relations
\begin{align}
    z&=\frac{1}{P}\sqrt{\frac{E^2_{\gamma^\prime}-m^2_{\gamma^\prime}}{1+\tan^2\theta}},\\
    k^2_\perp&=\frac{E^2_{\gamma^\prime}-m^2_{\gamma^\prime}}{1+\cot^2\theta},
\end{align}
and the consequent Jacobian
\begin{equation}
    J(E_{\gamma^\prime}, \theta)\equiv \frac{2E_{\gamma^\prime}}{P}\sqrt{E^2_{\gamma^\prime} - m^2_{\gamma^\prime}}\sin\theta. \label{eq:Jacobian}
\end{equation}

\begin{figure}[!htb]
	\begin{center}
		\begin{subfigure}{0.495\textwidth}
			\centering
                \includegraphics[width=\textwidth]{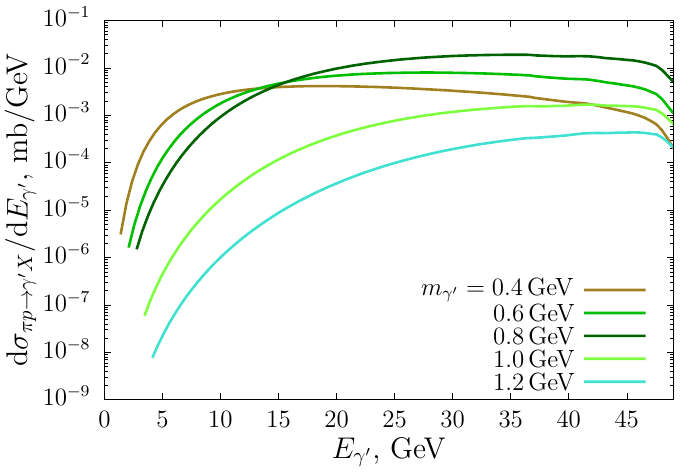}
			\caption{}
                \label{fig:spectra-brem}
		\end{subfigure}
        \hfill
		\begin{subfigure}{0.495\textwidth}
			\centering
			\includegraphics[width=\textwidth]{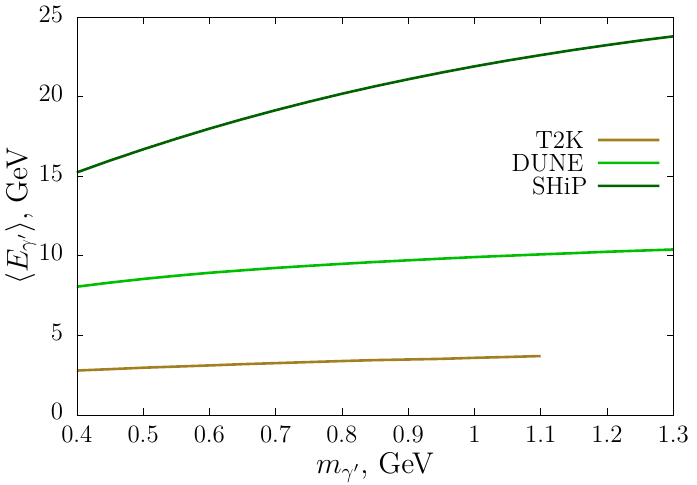}
			\caption{}
                \label{fig:meanE-brem}
		\end{subfigure}
	\end{center}
	\caption{(a) Differential cross section of inelastic pion bremsstrahlung~\eqref{eq:dphspectrum} for a set of dark photon masses, $m_{\gamma^\prime}$, listed in the legend, as a function of dark photon energy $E_{\gamma^\prime}$. The incident pion momentum $P=50\text{\,GeV}$ corresponds to NA64h experiment. (b) Mean energy of dark photons~\eqref{eq:meanE-brem} produced via inelastic pion bremsstrahlung as a function of dark photon mass for the incident pion momenta~\eqref{eq:Peff} typical for T2K, DUNE and SHiP experiments, see main text for details.}
\end{figure}

Combining~\eqref{eq:crsecfact} and \eqref{eq:Jacobian} brings us to the spectra of dark photons produced in the inelastic pion bremsstrahlung
\begin{equation} \label{eq:dphspectrum}
    \frac{\dd \sigma_{\pi p\rightarrow \gamma^\prime X}}{\dd E_{\gamma^\prime}} = \int_0^{\pi/2} \frac{\dd^2 \sigma_{\pi p\rightarrow \gamma^\prime X}}{\dd z \dd k^2_\perp} H_\text{int}\left(z, k^2_\perp\right)\Bigg\vert_{\raisebox{5pt}{$^{z=z(E_{\gamma^\prime}, \theta),}_{k_{\perp}=k_{\perp}(E_{\gamma^\prime}, \theta)}$}} J(E_{\gamma^\prime}, \theta) \dd \theta,
\end{equation}
where, satisfying the constraints imposed above \eqref{con-1}, \eqref{con-2}, \eqref{con-3}, we limit the region of integration by the product of Heaviside step functions $\Theta(x)$, 
\begin{align}
    H_\text{int}\left(z, k^2_\perp\right)&\equiv H_1\left(z, k^2_\perp\right)H_2\left(z, k^2_\perp\right)H_3\left(z, k^2_\perp\right),\\ 
    H_1\left(z, k^2_\perp\right)&\equiv \Theta\left(0.1-\frac{m^2_{\gamma^\prime}+k^2_\perp}{z^2P^2}\right), \label{eq:H1}\\
    H_2\left(z, k^2_\perp\right)&\equiv \Theta\left(0.1-\frac{m^2_\pi+k^2_\perp}{\left(1-z\right)^2P^2}\right),\\
    H_3\left(z, k^2_\perp\right)&\equiv \Theta\left(0.1-\frac{H_\pi}{4z\left(1-z\right)^2P^2}\right). \label{eq:H3}
\end{align}
We include in the spectrum~\eqref{eq:dphspectrum} all dark photons that are produced in the forward direction, assuming that the detector has the NA64h-like geometry.
Figure~\ref{fig:spectra-brem}
shows the spectra of dark photons of various masses $m_{\gamma^\prime}$ produced by inelastic pion bremsstrahlung for the pion beam of the NA64h experiment with $P=50\text{\,GeV}$. An interesting feature of the obtained spectra is the significant contribution from high-energy dark photons, whose decay products are much easier to detect. Because of this, the search for dark photons with masses 0.4--1.2\,GeV at the NA64h experiment with negatively charged pion beam looks quite promising.

We would also like to comment on dark photon production via pion brems\-strah\-lung in the experiments with proton beams, such as T2K, DUNE, and SHiP, where secondary pions appear inside the target and dump. Strictly speaking, simulations are required to describe the secondary pions' production in these experiments. For simplified estimates, we interpolate between the multiplicities of charged pion production $\left<\pi_\text{ch}(P_j)\right>$ obtained for different proton beam momenta $P_j$ in bubble chambers~\cite{Bonn-Hamburg-Munich:1974cjj,Morse:1976tu} and in CERN ISR~\cite{Ames-Bologna-CERN-Dortmund-Heidelberg-Warsaw:1983cqw}. The interpolation results for charged pion multiplicities $\left<\pi_\text{ch}(P)\right>$ at the momenta $P$ corresponding to J-PARC, Fermilab and CERN SPS proton beams are listed in table~\ref{tab:pimult}.
\begin{table}
    \centering
    \begin{tabular}{|c|c|c|c|c|}
        \hline
         & $P$,\,GeV & $\left<\pi_\text{ch}(P)\right>$ & $\theta_l$ & $\theta_u$\\
        \hline
        T2K & 30 & 4.5 & 0.030 & 0.040\\
        \hline
        DUNE & 120 & 6.6 & 0 & 0.0061\\
        \hline
        SHiP & 400 & 8.8 & 0 & 0.055\\
        \hline
    \end{tabular}
    \caption{Charged pion multiplicities $\left<\pi_\text{ch}(P)\right>$ at proton beam momenta $P$ corresponding to the T2K, DUNE, and SHiP experiments obtained by interpolating the results of bubble chambers and CERN ISR~\cite{Bonn-Hamburg-Munich:1974cjj,Morse:1976tu,Ames-Bologna-CERN-Dortmund-Heidelberg-Warsaw:1983cqw} and the limiting polar angles $\theta_l$ and $\theta_u$ at which the dark photon can reach the detectors of the listed experiments.}
    \label{tab:pimult}
\end{table}
Using charged pion multiplicities, one can express the effective mean momentum of the secondary pions as 
\begin{equation} \label{eq:Peff}
    P_\text{eff}\equiv\frac{2P}{3\left<\pi_\text{ch}(P)\right>},
\end{equation}
where we assume that neutral pions are produced with the multiplicity $\left<\pi_\text{0}(P)\right>\simeq \left<\pi_\text{ch}(P)\right>/2$ (see~\cite{HERMES:2001ghm} for its measurements in deep inelastic scattering in the HERMES experiment). For simplicity, we assume that most of the secondary pions are produced with the momenta directed along the initial proton beam. For every experiment, we estimate the smallest $\theta_l$ and the largest $\theta_u$ polar angles at which dark photon can reach the detector according to their dimensions presented in table~2 of~\cite{Gorbunov:2024vrc}. Their numerical values can be found in table~\ref{tab:pimult}. Next, similar to~\eqref{eq:dphspectrum} we integrate the differential cross section over polar angle from $\theta_l$ to $\theta_u$ to obtain the spectrum of dark photons $\dd \sigma_{\pi p\rightarrow \gamma^\prime X}/\dd E_{\gamma^\prime}$.
This allows us to estimate the full production cross section via inelastic pion bremsstrahlung 
\begin{equation} \label{eq:fullcrsec}
    \sigma_{\pi p\rightarrow \gamma^\prime X} = \int_{m_{\gamma^\prime}}^{P_\text{eff}} \frac{\dd \sigma_{\pi p\rightarrow \gamma^\prime X}}{\dd E_{\gamma^\prime}} \dd E_{\gamma^\prime}
\end{equation}
and the mean energy of dark photons  
\begin{equation} \label{eq:meanE-brem}
    \left<E_{\gamma^\prime}\right> = \frac{1}{\sigma_{\pi p\rightarrow \gamma^\prime X}} \int_{m_{\gamma^\prime}}^{P_\text{eff}} E_{\gamma^\prime} \frac{\dd \sigma_{\pi p\rightarrow \gamma^\prime X}}{\dd E_{\gamma^\prime}} \dd E_{\gamma^\prime}
\end{equation}
in the case of secondary pion scatterings. Figure~\ref{fig:meanE-brem} shows the dependence of the mean dark photon energy $\left<E_{\gamma^\prime}\right>$ on its mass $m_{\gamma^\prime}$ for T2K, DUNE, and SHiP experiments. The line corresponding to T2K experiment ends at the dark photon mass $m_{\gamma^\prime}$ about 1.1 GeV, since for heavier dark photons it is impossible to meet the conditions \eqref{eq:H1}--\eqref{eq:H3} simultaneously, which means that quasi-real approximation cannot describe the process of pion bremsstrahlung.

\begin{figure}
    \centering
    \includegraphics[width=0.6\linewidth]{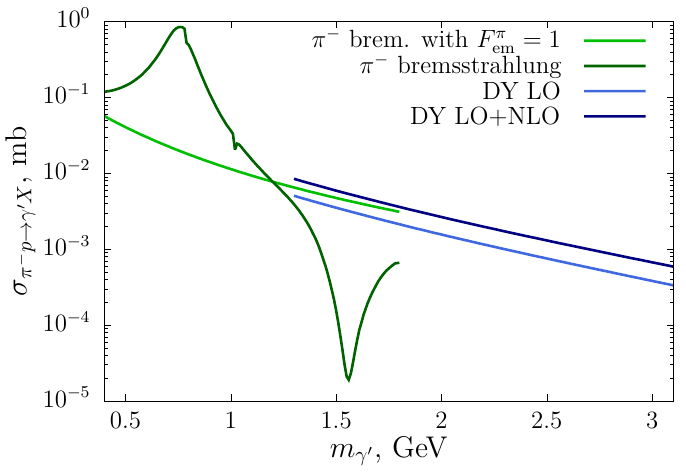}
    \caption{In green: the dependence of full inelastic pion bremsstrahlung cross section~\eqref{eq:fullcrsec} on dark photon mass $m_{\gamma^\prime}$ calculated with the pion electromagnetic form factor $F^\pi_\text{em}(m^2_{\gamma^\prime})$~\cite{RuizArriola:2024gwb} (dark green line) and with the same form factor put equal to unity (light green line). In blue: full cross section of dark photon production via Drell--Yan process in $\pi^-p\rightarrow\gamma^\prime X$ calculated within perturbative QCD. 
    Light blue line shows the LO result, whereas dark blue line presents the sum of LO and NLO cross sections. In all cases the calculations are made for the negatively charged pion beam of the NA64h experiment with $P=50\text{\,GeV}$.}
    \label{fig:pipcomp}
\end{figure}

Finally, in figure~\ref{fig:pipcomp}
the dark green line shows the dependence of the full inelastic pion bremsstrahlung cross section on dark photon mass $m_{\gamma^\prime}$ for NA64h pion beam with $P=50\text{\,GeV}$. A peak and a dip in the cross section appear due to the special behaviour of the pion electromagnetic form factor $F^\pi_\text{em}(m^2_{\gamma^\prime})$. According to the Gounaris-Sakurai model~\cite{Gounaris:1968mw}, the peak is caused by possible photon mixing with resonance $\rho(770)$, while the dip comes from destructive interference of $\rho^\prime(1450)$ and $\rho^{\prime\prime}(1700)$ resonance contributions. For comparison, the light green line in figure~\ref{fig:pipcomp} shows the full pion bremsstrahlung cross section with unity pion electromagnetic form factor. The blue lines in the same figure show the results obtained in the framework of perturbative QCD for the Drell--Yan-like process in $\pi^-p$-collisions, that is discussed in section~\ref{sec:qcd} below. We conclude that for fixed-target experiments with pion beam inelastic pion bremsstrahlung gives considerable input to the production of dark photons of masses 0.4--1.3\,GeV. 
\section{Drell--Yan-like process with dark photon in perturbative QCD} \label{sec:qcd}

Below we consider the dark photon production in the competitive Drell--Yan-like process in the framework of perturbative QCD. At the leading order (LO) in the strong coupling constant $\alpha_s$ on the parton level, this process is just the quark-antiquark annihilation with the subsequent dark photon decay $q\bar{q}\rightarrow \gamma^{\prime}\rightarrow e^+ e^-$. 

In order to compare our results with the cross section of dark photon production via inelastic pion bremsstrahlung~\eqref{eq:crsecfact}, we derive the analogous QCD cross section in the narrow-width approximation (NWA), which gives, for instance, at the LO on the parton level 
\begin{equation} \label{eq:NWA}
    \sigma_{q\bar{q}\rightarrow\gamma^\prime}=\sigma_{q\bar{q}\rightarrow\gamma^\prime\rightarrow e^+ e^-}\frac{ \Gamma_\text{tot}}{\Gamma_\text{lep}(m_e)},
\end{equation}
where $\Gamma_\text{tot}$ is the total decay width, that is the sum of dark photon decay widths to pairs of charged SM particles: electrons, muons, $\tau$-leptons and hadrons,
\begin{equation}
    \Gamma_\text{tot} = \Gamma_\text{lep}(m_e)+\Gamma_\text{lep}(m_\mu)+\Gamma_\text{lep}(m_\tau)+\Gamma_\text{had},
\end{equation}
the dark photon decay width to a charged lepton with mass $m_l$ is given by
\begin{equation}
    \Gamma_\text{lep}(m_l) = \frac{\epsilon^2 \alpha_\text{em}}{3}m_{\gamma^\prime}\sqrt{1-\frac{4m^2_l}{m^2_{\gamma^\prime}}}\left(1+\frac{2m^2_l}{m^2_{\gamma^\prime}}\right)
\end{equation}
and the dark photon decay width to hadrons is fixed by the experimentally measured $R$-ratio~\cite{ParticleDataGroup:2022pth}
\begin{equation}
    \Gamma_\text{had} = \Gamma_\text{lep}(m_\mu)R(m_{\gamma^\prime}).
\end{equation}

The Drell--Yan cross section is traditionally parametrized by two Lorentz-in\-va\-ri\-ant variables: $M^2$ which is the squared invariant mass of the $e^+e^-$-pair in the final state and the Feynman variable $x_F\equiv x_1-x_2$, where $x_1$ and $x_2$ are the light-cone momentum fractions of colliding partons~\cite{Golec-Biernat:2010dup}. In the LO these momentum fractions are related by the energy-momentum conservation, giving $x_1 x_2 = M^2/s \equiv \tau$.
Thus it is straightforward to express the momentum fractions in the LO in terms of variables $M^2$ and $x_F$ as follows
\begin{equation}
    x_{1,2}=\frac{1}{2}\left(\sqrt{x^2_F+4\tau}\pm x_F\right).
\end{equation}
In our calculation, we associate $x_1$ with a parton from the negatively charged pion and $x_2$ with a parton from the target proton.

Next, we modify the standard expression for the cross section of the Drell--Yan process\,\cite{Golec-Biernat:2010dup} by replacing a factor $1/M^4$ from the squared massless SM photon propagator with the squared Breit--Wigner dark photon propagator
\begin{equation} \label{eq:BW}
    |\mathcal{P}_\text{BW}|^2=\frac{1}{(M^2-m^2_{\gamma^\prime})^2+m^2_{\gamma^\prime} \Gamma^2_\text{tot}}.
\end{equation}
At the LO, this results in the following differential cross section for the Drell--Yan-like process with the dark photon
\begin{equation}
    \begin{split}
        \frac{\dd^2 \sigma_\text{LO}}{\dd M^2 \dd x_F}=&\frac{4\pi \alpha^2_\text{em}}{9} \frac{x_1x_2}{x_1+x_2} |\mathcal{P}_\text{BW}|^2 \times \\ & \times \sum_{f=u,d} e^2_f \left(q_f(x_1, M^2)\bar{q}_f(x_2, M^2) + \bar{q}_f(x_1, M^2)q_f(x_2, M^2)\right),
    \end{split}
\end{equation}
where $e_f$ is the quark electric charge, $q_f(x_i, M^2)$ and $\bar{q}_f(x_i, M^2)$ are the quark and antiquark distribution functions computed for the momentum fraction $x_i$ at the factorization scale $M^2$. We also modify the NLO correction to the Drell--Yan-like cross section in the $\overline{\text{MS}}$ scheme~\cite{Golec-Biernat:2010dup,Altarelli:1978id,Altarelli:1979ub,Kubar-Andre:1978eri} in the same way by changing the SM photon propagator squared to~\eqref{eq:BW}.

Our final differential cross section of the Drell--Yan-like process with the dark photon is the sum of the LO and the NLO contributions
\begin{equation} \label{eq:diffDY}
    \frac{\dd^2 \sigma_\text{DY}}{\dd M^2 \dd x_F} = \frac{\dd^2 \sigma_\text{LO}}{\dd M^2 \dd x_F} +  \frac{\dd^2 \sigma_\text{NLO}}{\dd M^2 \dd x_F}.
\end{equation}
To obtain the full Drell--Yan-like process cross section, we integrate it over the squared invariant mass $M^2$ of $e^+e^-$-pair and over the Feynman variable $x_F$
\begin{equation} \label{eq:DY}
    \sigma_\text{DY} = \int_{M^2_-}^{M^2_+} \dd M^2 \int_{-1+\tau}^{1-\tau} \dd x_F \frac{\dd^2 \sigma_\text{DY}}{\dd M^2 \dd x_F},
\end{equation}
where the limits of integration over $M^2$ are $M^2_\pm=m^2_{\gamma^\prime}\pm 3m_{\gamma^\prime}\Gamma_\text{tot}$. Finally, in the NWA~\eqref{eq:NWA} we obtain from the Drell--Yan-like cross section~\eqref{eq:DY} the cross section of dark photon production $\pi^-p\rightarrow\gamma^\prime X$. The numerical calculations are then performed with the help of the \texttt{LHAPDF} library~\cite{Buckley:2014ana} and parton distribution functions set \texttt{JAM21PionPDFnlo}~\cite{Barry:2021osv} for charged pion and \texttt{CT14lo}, \texttt{CT14nlo} sets~\cite{Dulat:2015mca} for proton. 

In order to compare the results of the section~\ref{sec:splitting} with the result for the Drell--Yan-like process, we plot full cross sections of the inelastic process of dark photon production in $\pi^-p$-collisions in figure~\ref{fig:pipcomp} in green (pion bremsstrahlung, see section~\ref{sec:splitting} for details) and in blue (competitive QCD process). There the light blue line shows the full cross section of dark photon production, calculated at the LO in the strong coupling constant $\alpha_s$ for the negatively charged pion beam with the momentum $P=50\text{ GeV}$ that is used in the NA64h experiment. The dark blue line in the same figure shows the sum of the LO and the NLO contributions to the same cross section. One can see that for dark photon masses $m_{\gamma^\prime}$ greater than 1.3\,GeV the Drell--Yan-like process clearly makes the dominant contribution to dark photon production. 

Next, we would like to study the energy distribution of dark photons produced in the Drell--Yan-like process. At first, we express the momentum fractions in terms of hadrons momenta by their definition~\cite{Meyer-Conde:2019frd}
\begin{equation}
    x_1 \equiv \frac{M^2}{2(p\cdot k)},\quad 
    x_2 \equiv \frac{M^2}{2(P_i\cdot k)}.
\end{equation}
This gives us the Feynman variable as a function of the squared invariant mass $M^2$, the dark photon energy $E_{\gamma^\prime}$, and the dark photon squared transverse momentum $k^2_\perp$,
\begin{equation} \label{eq:xF}
    x_F(M^2, E_{\gamma^\prime}, k^2_\perp) = \frac{M^2}{2E_\pi\left(E_{\gamma^\prime}-\sqrt{E^2_{\gamma^\prime}-M^2-k^2_\perp}\right)} - \frac{M^2}{2M_p E_{\gamma^\prime}}.
\end{equation}
To obtain the dark photon energy distribution, below we change the Feynman variable $x_F$ to the dark photon energy $E_{\gamma^\prime}$ with the Jacobian
\begin{equation} \label{eq:xFJac}
    \pdv{x_F}{E_{\gamma^\prime}} = \frac{M^2}{2E_\pi}\cdot \frac{1}{\sqrt{E^2_{\gamma^\prime}-M^2-k^2_\perp}\left(E_{\gamma^\prime}-\sqrt{E^2_{\gamma^\prime}-M^2-k^2_\perp}\right)} + \frac{M^2}{2M_p E^2_{\gamma^\prime}}.
\end{equation}

Note that the expressions~\eqref{eq:xF},\,\eqref{eq:xFJac} explicitly depend on the square of the transverse momentum $k^2_\perp$ of the dark photon or, equivalently, of the lepton pair in the final state. For the SM Drell--Yan process $\pi^-N\rightarrow \mu^+\mu^-X$, the average absolute value of transverse momentum $\langle k_\perp\rangle$ has been measured by OMEGA, E537, NA3, and CIP experiments at different center-of-mass energies $\sqrt{s}$. To estimate the value of $\langle k_\perp\rangle$ for the purposes of this work, we implement the numerical fit~\cite{Malhotra:1982cx}
\begin{equation}
    \langle k_\perp\rangle = 0.48\text{\,GeV} + 0.034\sqrt{s}.
\end{equation}
Besides, the experimentally measured distribution of the differential Drell--Yan cross section over the square of lepton pair transverse momentum $k^2_\perp$ at fixed variables $M^2$ and $x_F$ has the form~(see  subsection 27.3 of \cite{Meyer-Conde:2019frd})
\begin{equation} \label{eq:kTdistr}
    \frac{\dd^3 \sigma_\text{DY}}{\dd M^2 \dd x_F \dd k^2_\perp} = a \left(1+\frac{k^2_\perp}{b^2}\right)^{-6},
\end{equation}
where the normalization factor $a$ can be restored using the differential Drell--Yan process cross section analogous to~\eqref{eq:diffDY},
\begin{equation}
    a = \frac{\dd^2\sigma_\text{DY}}{\dd M^2 \dd x_F} \cdot \frac{5}{b^2}.
\end{equation}
By calculating the average absolute value of the transverse dark photon momentum $\langle k_\perp\rangle$ from the distribution~\eqref{eq:kTdistr}, it is possible to restore the value of the second parameter
\begin{equation} \label{eq:b}
    b = \frac{256}{35\pi} \langle k_\perp\rangle.
\end{equation}

\begin{figure}[!htb]
	\begin{center}
		\begin{subfigure}{0.495\textwidth}
			\centering
                \includegraphics[width=\textwidth]{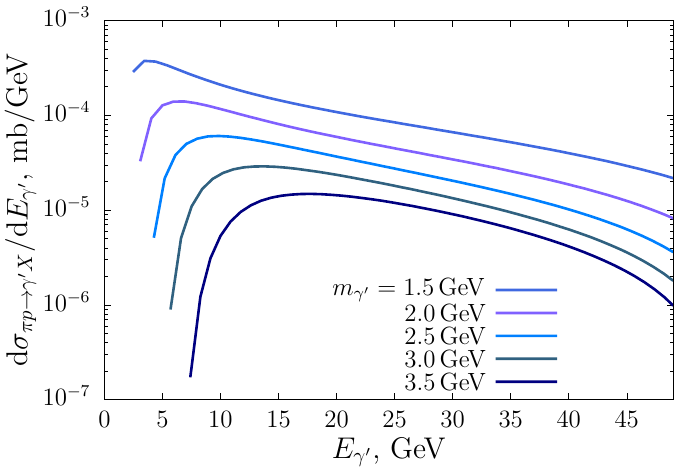}
			\caption{}
                \label{fig:spectra-qcd}
		\end{subfigure}
        \hfill
		\begin{subfigure}{0.495\textwidth}
			\centering
			\includegraphics[width=\textwidth]{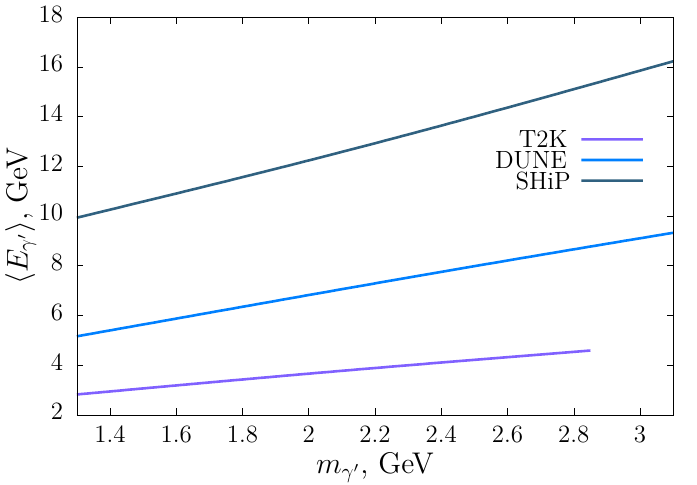}
			\caption{}
                \label{fig:meanE-qcd}
		\end{subfigure}
	\end{center}
	\caption{(a) Differential cross section~\eqref{eq:dphEdistr} of dark photon production in the Drell--Yan-like process in the NWA~\eqref{eq:NWA} for a set of dark photon masses, $m_{\gamma^\prime}$, listed in the legend, as a function of dark photon energy $E_{\gamma^\prime}$. The incident pion momentum $P=50\text{\,GeV}$ corresponds to NA64h experiment. (b) Mean energy of dark photons~\eqref{eq:meanE-qcd} produced via Drell--Yan-like process as a function of dark photon mass for the incident pion momenta~\eqref{eq:Peff} typical for T2K, DUNE and SHiP experiments, see main text for details.}
\end{figure}

Combining \eqref{eq:xFJac}--\eqref{eq:b} allows us to obtain the dark photon energy spectrum upon integrating over the lepton pair masses squared $M^2$ and the transverse dark photon momenta squared $k^2_\perp$,
\begin{equation} \label{eq:dphEdistr}
    \frac{\dd \sigma_\text{DY}}{\dd E_{\gamma^\prime}} = \int_{M^2_-}^{M^2_+} \dd M^2 \int_0^{b^2} \dd k^2_\perp \abs{\frac{\partial x_F}{\partial E_{\gamma^\prime}}} \frac{\dd^2 \sigma_\text{DY}}{\dd M^2 \dd x_F} \frac{5}{b^2} \left(1+\frac{k^2_\perp}{b^2}\right)^{-6}.
\end{equation}
Figure~\ref{fig:spectra-qcd} 
shows the dependence of the differential cross section~\eqref{eq:dphEdistr} divided by the dark photon branching to $e^+e^-$ in the NWA~\eqref{eq:NWA} on the dark photon energy $E_{\gamma^\prime}$ for a set of dark photon masses $m_{\gamma^\prime}$ in the interval 1.5--3.5\,GeV and for the incident pion momentum $P=50\text{\,GeV}$. In contrast to inelastic bremsstrahlung, in this case the maximum of the dark photon spectrum is at energies $E_{\gamma^\prime}$ lower than $P/2$. 

This feature is important for NA64h setup, where the missing energy is utilized as the main signature of new particles. This feature can be potentially used in experiments that search for the vector decays. In models with vector mass of about 1.3\,GeV both mechanisms give similar contributions to the vector production. Interestingly, they can be distinguished by measuring the average energies of the signal events.

It seems that for a dark photon with known energy and intermediate mass $m_{\gamma^\prime}$ of about 1.3\,GeV, for which both pion bremsstrahlung and Drell--Yan-like processes can potentially contribute to production (see figure~\ref{fig:pipcomp}), these qualitative differences allow us to determine the most probable mechanism.

Next, to characterize the dark photon production from secondary pions with momentum $P_\text{eff}$, one can calculate the dark photon energy averaged over the distribution~\eqref{eq:dphEdistr}
\begin{equation} \label{eq:meanE-qcd}
    \langle E_{\gamma^\prime}\rangle=\frac{1}{\sigma_\text{DY}} \int^{P_\text{eff}}_{m_{\gamma^\prime}}E_{\gamma^\prime} \frac{\dd \sigma_\text{DY}}{\dd E_{\gamma^\prime}} \dd E_{\gamma^\prime}.
\end{equation}
The mean energies of dark photons produced in the Drell--Yan-like process as functions of dark photon mass $m_{\gamma^\prime}$ are presented in figure~\ref{fig:meanE-qcd} for the secondary pions from T2K, DUNE and SHiP experiments with the effective incident momenta~\eqref{eq:Peff}.
\section{Discussion} \label{sec:discussion}

To summarize, we have revisited the dark photon production in the negatively charged pion-proton collisions $\pi^-p\rightarrow \gamma^\prime X$. We have demonstrated the inapplicability of LO ChPT, originally implemented to this problem in~\cite{Curtin:2023bcf}, if the value of dark photon mass $m_{\gamma^\prime}$ is $\mathcal{O}(1)$\,GeV and the pion beam momentum $P\sim50$\,GeV (as it is in the NA64h experiment). Despite this, we expect that elastic pion bremsstrahlung still can be described using the diffractive physics methods, e.g. in Regge regime of the holographic QCD model~\cite{Liu:2023tjr}. We have proposed an alternative way to estimate the cross section of inelastic pion bremsstrahlung $\pi^-p\rightarrow\gamma^\prime X$ with the use of the factorization procedure~\eqref{eq:crsecfact}, and we have derived the new splitting function $w_\pi(z, k^2_\perp)$ for the charged pion~\eqref{eq:splitfunc}. Moreover, we have also considered the competing perturbative QCD process of Drell--Yan type, where one of the partons is coming from $\pi^-$, and the other one from $p$ and have numerically estimated its cross section in both LO and NLO. 

The comparison of the total cross sections for the negatively charged pion beam momentum $P=50$\,GeV corresponding to the NA64h experiment (see figure~\ref{fig:pipcomp}) shows that due to the amplification by the pion electromagnetic form factor, the inelastic pion bremsstrahlung process makes the main contribution to the production of dark photons with masses $m_{\gamma^\prime}=$\,0.4--1.3\,GeV. However, at $m_{\gamma^\prime}=$\,1.3--3.5\,GeV the Drell--Yan-like QCD process dominates. Bear in mind that the earlier study~\cite{Curtin:2023bcf} considered only the proton-induced Drell--Yan-like process and concluded that in the SpinQuest experiment its input is suppressed in comparison with the pion bremsstrahlung (estimated using LO ChPT beyond the limits of its applicability). It is worth reevaluating the prospects of SpinQuest and its successor DarkQuest in the search for dark photons.

Finally, we have studied the energy distributions of the dark photons produced both in the inelastic pion bremsstrahlung~\eqref{eq:dphspectrum} and in the Drell--Yan-like process~\eqref{eq:dphEdistr}. We conclude that a noticeable fraction of high-energetic dark photons from pion bremsstrahlung makes them detectable (if any are produced) in the NA64h experiment. We have also estimated the mean energies~\eqref{eq:meanE-brem}, \eqref{eq:meanE-qcd} of the dark photons in the mass range $m_{\gamma^\prime}=$\,0.4--3.5\,GeV produced from the secondary pions with the momenta typical for the T2K, DUNE and SHiP experiments, since pion bremsstrahlung is also the novel dark photon production mechanism in proton beam-dump experiments.

\acknowledgments

We would like to thank S.~Gninenko and D.~Kirpichnikov for valuable discussions on NA64h prospects. We are also grateful to S.~Demidov, A.~Kataev, V.~Kim, M.~Kri\-vo\-ru\-chen\-ko and P.~Reimitz for interesting comments. The numerical study of dark photon production via Drell--Yan-like process is supported by the Russian Science Foundation grant No. 25-12-00309. The work of EK on the factorization of inelastic pion bremsstrahlung is supported by the PhD fellowship No. 21-2-10-37-1 of the Foundation for the Advancement of Theoretical Physics and Mathematics “BASIS”.

\bibliographystyle{JHEP}
\bibliography{refs.bib}

\end{document}